\documentclass[prb,twocolumn,showpacs]{revtex4}     
\usepackage{epsfig,amsmath} 

\begin{document}

\title{Weak localization and magnetoconductance of Dirac fermions under charged impurities in graphene} 

\author{Xin-Zhong Yan$^{1}$ and C. S. Ting$^2$}
\affiliation{$^{1}$Institute of Physics, Chinese Academy of Sciences, P.O. Box 603, 
Beijing 100190, China\\
$^{2}$Texas Center for Superconductivity, University of Houston, Houston, Texas 77204, USA}
 
\date{\today}
 
\begin{abstract}
On the basis of self-consistent Born approximation, we present a theory of weak localization of Dirac fermions under finite-range scatters in graphene. With an explicit solution to the ground state of singlet pseudospin Cooperons, we solve the Bethe-Salpeter matrix equation for all the singlet and triplet pseudospin Cooperons at long-wave length states by perturbation treatment. The solution to the Cooperon in the presence of the external weak magnetic field is also obtained. We calculate the quantum interference correction to the conductivity and present the comparison with experiments. It is shown that the present calculation for the magnetoconductivity is in good agreement with some of the experimental measurements.
\end{abstract}

\pacs{73.20.Fz, 72.10.Bg, 73.50.-h, 81.05.Uw} 

\maketitle

\section{Introduction}

It has been found that the charged impurities with screened Coulomb potentials \cite{Nomura,Hwang,Yan} are responsible for the observed carrier density dependence of the electric conductivity of graphene.\cite{Geim} In a recent work, we have investigated the weak localization (WL) of electrons under the charged impurity scattering in graphene.\cite{Yan2} The description for the Cooperons under the finite-range scatters is different from that for the zero-range potentials as studied in the existing works.\cite{Ziegler,Suzuura,Khveshchenko,McCann} In this paper, we present the details of the formalism for the WL of Dirac fermions under finite-range scatters in graphene. We also calculate the quantum inetrference correction (QIC) to the electric conductivity under a weak magnetic field and compare the result for the magnetoconductivity with the experimental measurements.  

The central problem of theoretically studying the weak localization of Dirac fermions under finite-range scatters is to solve the Bethe-Salpeter matrix equation for the Cooperons. With the self-consistent Born approximation (SCBA) to the single particle, we can obtain an explicit solution to the ground state of the singlet pseudospin Cooperons. By perturbation method, we will solve the Bethe-Salpeter matrix equation for all the singlet and triplet pseudospin Cooperons at long-wave length states. With the Cooperons, we derive the quantum interference correction to the electric conductivity that gives rise to WL effect.

At low carrier doping, the low energy excitations of electrons in graphene can be viewed as massless Dirac fermions.\cite{Wallace,Ando,Castro,McCann1,Yan1} This has been confirmed by recent experiments. \cite{Geim,Zhang} Using the Pauli matrices $\sigma$'s and $\tau$'s to coordinate the electrons in the two sublattices ($a$ and $b$) of the honeycomb lattice and two valleys (1 and 2) in the first Brillouin zone, respectively, and suppressing the spin indices for briefness, the Hamiltonian of the system is given by
\begin{equation}
H = \sum_{k}\psi^{\dagger}_{k}v\vec
 k\cdot\vec\sigma\tau_z\psi_{k}+\frac{1}{V}\sum_{kq}\psi^{\dagger}_{k-q}V_i(q)\psi_{k} \label{H}
\end{equation}
where $\psi^{\dagger}_{k}=(c^{\dagger}_{ka1},c^{\dagger}_{kb1},c^{\dagger}_{kb2},c^{\dagger}_{ka2})$ is the fermion operator, the momentum $k$ is measured from the center of each valley, $v$ ($\sim$ 5.856 eV\AA) is the velocity of electrons, $V$ is the volume of system, and $V_i(q)$ is the finite-range impurity potential. For charged scatters, $V_i(q)$ is given by,    
\begin{equation}
V_i(q) = 
\begin{pmatrix}
n_i(-q)v_0(q)\sigma_0& n_i(Q-q)v_1\sigma_1 \\
n_i(-Q-q)v_1\sigma_1& n_i(-q)v_0(q)\sigma_0 
\end{pmatrix}\label{vi}
\end{equation}
where $n_i(-q)$ is the Fourier component of the impurity density, $v_0(q)$ and $v_1$ are respectively the intravalley and intervalley impurity scattering potentials, and $Q$ is a vector from the center of valley 2 to that of the valley 1 [Fig. 1(a)]. In Appendix, we detail the discussion on this impurity potential. Here, all the momenta are understood as vectors. 

Under the SCBA [Fig. 1(b)],\cite{Gorkov,Lee1,Fradkin} the Green function $G(k,\omega)=[\omega+\mu-v\vec
 k\cdot\vec\sigma\tau_z-\Sigma(k,\omega)]^{-1}$ and the self-energy $\Sigma(k,\omega)=\Sigma_0(k,\omega)+\Sigma_c(k,\omega)\hat k\cdot\vec\sigma\tau_z$ of the single particles are determined by coupled integral equations:\cite{Yan} 
\begin{eqnarray}
\Sigma_0(k,\omega) &=& \frac{n_i}{V}\sum_{k'}[v^2_0(|k- k'|)+v^2_1]G_0(k',\omega)\label{sc2}\\ 
\Sigma_c(k,\omega) &=& \frac{n_i}{V}\sum_{k'}v^2_0(|k-k'|)
G_c(k',\omega)\hat k\cdot\hat k'\label{sc3}
\end{eqnarray}
with 
\begin{eqnarray}
G_0(k,\omega) &=& \frac{\tilde\omega}{\tilde\omega^2-h_{k}^2}, \nonumber\\
G_c(k,\omega) &=& \frac{h_{k}}{\tilde\omega^2-h_{k}^2} \nonumber
\end{eqnarray}
where $\tilde\omega=\omega+\mu-\Sigma_0(k,\omega)$ with $\mu$ the chemical potential, $h_k = vk+\Sigma_c(k,\omega)$, $\hat k$ is the unit vector in $k$ direction, and the frequency $\omega$ is understood as a complex quantity with infinitesimal small imaginary part. The current vertex $v\Gamma_x(k,\omega_1,\omega_2)$ [Fig. 1(c)] can be expanded as
\begin{equation}
\Gamma_x(k,\omega_1,\omega_2)=\sum_{j=0}^3y_j(k,\omega_1,\omega_2)A^x_j(\hat k) \label{vt}
\end{equation}
where $A^x_0(\hat k)=\tau_z\sigma_x$, $A^x_1(\hat k)=\sigma_x\vec\sigma\cdot\hat k$, $A^x_2(\hat k)=\vec\sigma\cdot\hat k\sigma_x$, $A^x_3(\hat k)=\tau_z\vec\sigma\cdot\hat k\sigma_x\vec\sigma\cdot\hat k$, and $y_j(k,\omega_1,\omega_2)$ are determined by four-coupled integral equations.\cite{Yan} The $x$-direction current-current correlation function [Fig. 2(a)] is obtained as
\begin{eqnarray}
P(\omega_1,\omega_2)
= \frac{2v^2}{V}\sum_{kj}y_j(k,\omega_1,\omega_2)X_j(k,\omega_1,\omega_2)\nonumber
\end{eqnarray}
with $X_j(k,\omega_1,\omega_2) = {\rm Tr}[G(k,\omega_1)A^x_j(\hat k)G(k,\omega_2)A^x_0(\hat k)]$, for $\omega$'s ($\omega_1$ and $\omega_2$) = $\omega\pm i0 \equiv \omega^{\pm}$. The detailed derivations of  
$\Gamma_x(k,\omega_1,\omega_2)$ and $P(\omega_1,\omega_2)$ can be found in Ref. \onlinecite{Yan}, and will not be repeated here.

\begin{figure} 
\centerline{\epsfig{file=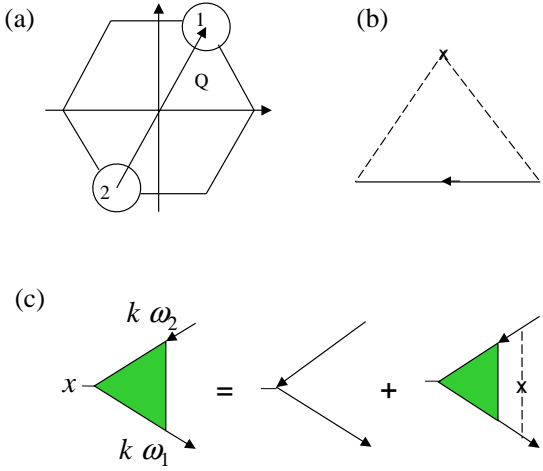,width=8. cm}}
\caption{(color online) (a) Brillouin zone and the two Dirac-cone valleys. (b) Self-consistent Born approximation for the self-energy. The solid line with arrow is the Green function. The dashed line is the effective impurity potential. (c) Current vertex with impurity insertions.}\label{fig1}
\end{figure} 

\section{Formalism}

The WL effect in the electric conductivity stems from the QIC to the electric conductivity. Theoretically, it is given by the maximum crossing diagrams as shown by Fig. 2(b).\cite{Suzuura,Fradkin,Abrahams,Lee} The process of the maximum crossing diagrams is associated with two-particle propagator $C^{j_1j_2j_3j_4}_{\alpha_1\alpha_2\alpha_3\alpha_4}(k,k',q,\omega)$ (Cooperon). It obeys the Bethe-Salpeter $16\times 16$ matrix equation represented in Fig. 2(c). Here, the superscripts $j$'s denote the valley indices, and the subscripts $\alpha$'s correspond to the sublattice indices. To explicitly write out the equation of Fig. 2(c), we here give the simpler one for $\tilde C^{j_1j_2j_3j_4}_{\alpha_1\alpha_2\alpha_3\alpha_4}(k,k',q,\omega)$ that starts from the single impurity line [the dashed line with a cross in Fig. 2(c)], using the convention $\delta^{j_1j_2}_{\alpha_1\alpha_2} = \delta_{j_1j_2}\delta_{\alpha_1\alpha_2}$ and $\bar{j}$ ($\bar\alpha$) as the conjugate valley (site) of $j$ ($\alpha$):
\begin{widetext}
\begin{eqnarray}
\tilde C^{j_1j_2j_3j_4}_{\alpha_1\alpha_2\alpha_3\alpha_4}(k,k',q,\omega) &=&
n_iv^2_0(|k-k'|)\delta^{j_1j_3}_{\alpha_1\alpha_3}\delta^{j_2j_4}_{\alpha_2\alpha_4}
+n_iv^2_1\delta^{j_1\bar j_3}_{\alpha_1\bar\alpha_3}\delta^{\bar j_2j_4}_{\bar\alpha_2\alpha_4}\delta_{j_1\bar j_2} \nonumber\\
& &+ \frac{n_i}{V}\sum_{k_1\beta\beta'}v^2_0(|k-k_1|)G^{j_1j_1}_{\alpha_1\beta}(q/2+k_1,\omega^+)G^{j_2j_2}_{\alpha_2\beta'}(q/2-k_1,\omega^-)\tilde C^{j_1j_2j_3j_4}_{\beta\beta'\alpha_3\alpha_4}(k_1,k',q,\omega) \nonumber\\
& &+ \frac{n_i}{V}\sum_{k_1\beta\beta'}v^2_1G^{\bar j_1\bar j_1}_{\bar\alpha_1\beta}(q/2+k_1,\omega^+)G^{j_1j_1}_{\bar\alpha_2\beta'}(q/2-k_1,\omega^-)\tilde C^{\bar j_1j_1j_3j_4}_{\beta\beta'\alpha_3\alpha_4}(k_1,k',q,\omega)\delta_{j_1\bar j_2}. \label{cp0}
\end{eqnarray}
\end{widetext}
The first term in the first line in right hand side of Eq. (\ref{cp0}) is due to the intravalley scatterings, while the second term comes from the intervalley scatterings. $\delta_{j_1\bar j_2}$ means that when a particle is scattered to valley $j_1$, another particle should be scattered to valley $\bar j_1$ so that the total momentum (vanishing small under consideration) of the Cooperon is unchanged. The second and third lines are the processes of Cooperon propagating after the intravalley and intervalley scatterings, respectively. The equation for $C$ is obtained from Eq. (\ref{cp0}) by subtracting the single impurity line from $\tilde C$.

\begin{figure} 
\centerline{\epsfig{file=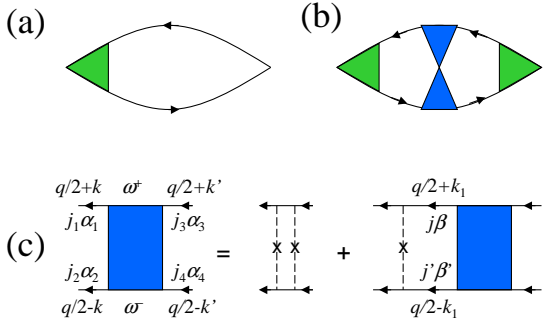,width=8. cm}}
\caption{(color online) (a) Electric conductivity. (b) Quantum interference correction to the conductivity. (c) Cooperon propagator.}
\end{figure} 

The form of Eq. (\ref{cp0}) seems rather miscellaneous. It may be simplified by classifying it with good quantum number of the Cooperons. To do this, we note that the elements of the coefficient matrix of $\tilde C$ in Eq. (\ref{cp0}) are arranged according to the indices (superscripts and subscripts) of the Green functions. Since the Green function $G(k,\omega)$ are composed by the unit matrix and $\tau_z\vec\sigma\cdot\vec k$, we then look for all the operators that commute with $\tau_z\vec\sigma\cdot\vec k$. Recently, McCann {\it et al.}\cite{McCann} have introduced the operators of isospin $\Sigma$'s and pseudospin $\Lambda$'s,
\begin{eqnarray}
\Sigma_0 &= \tau_0\sigma_0,~~~\Sigma_x = \tau_z\sigma_x,~~~\Sigma_y = \tau_z\sigma_y,~~~\Sigma_z = \tau_0\sigma_z, \nonumber\\
\Lambda_0 &= \tau_0\sigma_0,~~~\Lambda_x = \tau_x\sigma_z,~~~\Lambda_y = \tau_y\sigma_z,~~~\Lambda_z = \tau_z\sigma_0.\nonumber
\end{eqnarray}
Clearly, $\Lambda$'s commute with $\Sigma$'s and $\tau_z\vec\sigma\cdot\vec k$, and are conserving operations for the Cooperons. Therefore, we transform the Cooperons from the valley-sublattice space into the isospin-pseudospin space according to McCann {\it et al.},
\begin{equation}
C^{ll'}_{ss'} =
\frac{1}{4}\sum_{\{j,\alpha\}}(M^{l}_{s})^{j_1j_2}_{\alpha_1\alpha_2}C^{j_1j_2j_3j_4}_{\alpha_1\alpha_2\alpha_3\alpha_4}(M^{l'\dagger}_{s'})^{j_4j_3}_{\alpha_4\alpha_3} \label{vlpi}
\end{equation}
where $M^l_s = \Sigma_y\Sigma_s\Lambda_y\Lambda_l$. We will hereafter occasionally use the indices 0,1,2,3 or 0,x,y,z to label $l$ and $s$. In the isospin-pseudospin space, the single impurity line is given by $W^{ll'}_{ss'} = W^l_{s}\delta^{ll'}_{ss'}$ with
\begin{equation}
W^l_{s}(|k-k_1|) = n_iv^2_0(|k-k_1|)+n_iv^2_1(\delta_{l0}-\delta_{lz})(-1)^s, \nonumber
\end{equation}
which is the transform of the first line in the right hand side of Eq. (\ref{cp0}). The result of second+third lines in right hand side of Eq. {\ref{cp0}) is transformed to 
\begin{equation}
\frac{1}{V}\sum_{k_1,s_1} [W^l(|k- k_1|)\hat h(k_1,q)]_{ss_1}\tilde C^{ll'}_{s_1s'}(k_1,k',q) \nonumber
\end{equation}
where $\hat h(k_1,q)$ is a matrix defined in the isospin space with the element given by 
\begin{eqnarray}
h_{ss'}( k_1, q) = {\rm Tr}[G(-k^+_1,\omega^+)\Sigma_sG(-k^-_1,\omega^-)\Sigma^{\dagger}_{s'}]/4 \label{hkq}
\end{eqnarray}
and $k^{\pm}_1 = k_1\pm q/2$. With these results, we obtain the equation for $C^{ll'}_{ss'}$,
\begin{eqnarray}
C^{ll'}_{ss'}(k,k',q) &=& \frac{1}{V}\sum_{k_1,s_1}\Pi^l_{ss_1}(k,k_1,q)[W^l_{s_1}(|k_1-k'|)\delta^{ll'}_{s_1s'}\nonumber \\
& & ~~~~~~+C^{ll'}_{s_1s'}(k_1,k',q)]  \label{bsh0}
\end{eqnarray}
where $\hat \Pi^l(k,k_1,q) = \hat W^l(|k- k_1|)\hat h(k_1,q)$. Here, the argument $\omega$ of $C^{ll'}_{ss'}$ and $\Pi^l$ has been suppressed for briefness. From Eq. (\ref{bsh0}), it is seen that the pseudospin of the Cooperon is indeed conserved during the impurity scatterings, $C^{ll'}_{ss'}=C^{l}_{ss'}\delta_{ll'}$. We then need to deal with $C^l_{ss'}$. Thus, the original $16\times 16$ matrix equation is separated into four $4\times 4$ ones, each of them corresponding to a definite pseudospin $l$. In the isospin space, the Cooperon of a pseudospin $l$ is a $4\times 4$ matrix denoted as $C^l$. 

To solve Eq. (\ref{bsh0}), we use the standard method that expands $C^l$ in terms of the eigenfunctions $\Psi^l_n(k,q)$ of $\Pi^l(k,k_1,q)$: 
\begin{eqnarray}
C^l(k,k',q) = \sum_nc^l_n(q)\Psi^l_n(k,q) \Psi^{l\dagger}_n(k',q), \label{exp}
\end{eqnarray}
where $c^l_n(q)$ are constants and 
\begin{eqnarray}
\frac{1}{V}\sum_{k_1}\Pi^l(k,k_1,q) \Psi_{n}(k_1,q) = \lambda^l_n(q) \Psi_{n}(k,q) \label{egn1}
\end{eqnarray}
with $\lambda^l_n(q)$ the eigenvalue. Here, $\Psi^l_n(k,q)$ is a column vector with four components in the isospin space since $\Pi^l(k,k_1,q)$ is a $4\times 4$ matrix in this space. The constants $c^l_n(q)$ are determined by substituting Eq. (\ref{exp}) into Eq. (\ref{bsh0}). It is then seen that $c^l_n(q) \propto [1-\lambda^l_n(q)]^{-1}$. Therefore, the predominant contribution to $C^l$ comes from the state with the lowest $|1-\lambda^l_n(q)|\equiv |1-\lambda^l(q)|$ that can be vanishing small. We will here take into account only the state of the lowest $|1-\lambda^l(q)|$ for each $l$. 

Firstly, we consider the case of $l = 0$ and $q = 0$. A solution can be explicitly obtained as $\lambda^0(0) = 1$, and $\Psi^0(k,0)\equiv \Psi(k)$ with 
\begin{eqnarray}
\Psi^t(k) = [\Delta_0(k,\omega),-\Delta_c(k,\omega)\cos\phi,-\Delta_c(k,\omega)\sin\phi,0]\nonumber
\end{eqnarray}
where $\Psi^t(k)$ is the transpose of $\Psi(k)$, $\Delta_0(k,\omega)$ = Im$\Sigma_0(k,\omega^-)$, $\Delta_c(k,\omega)$ = Im$\Sigma_c(k,\omega^-)$, and $\phi$ is the angle of $k$. The four components of $\Psi(k)$ correspond to $s = 0, x, y,z$ respectively. The solution of $\lambda^0(0) = 1$ is the most important one which gives rise to the diverging contribution to the Cooperon. One may check this result with the help of Eqs. (\ref{sc2}) and (\ref{sc3}). Actually, the above solution is just a consequence of the Ward identity (under the SCBA for the self-energy): 
\begin{equation}
{\rm Im}\Sigma(k,\omega^-) = \frac{1}{V^2}\sum_{k'}\langle V_i(k-k'){\rm Im}G(k',\omega^-)V_i(k'-k)\rangle \nonumber
\end{equation}
where $\langle\cdots\rangle$ means the average over the impurity distributions [Fig. 1(b)]. There are three non-vanishing components in $\Psi(k)$ because of the finite-range impurity scatterings. For the zero-range potential, only the first component of $\Psi(k)$ survives and is a constant. One then needs to solve a scalar equation instead of the matrix integral equation.

For finite but small $q$, by expanding $\hat\Pi^0(k,k',q)$ to second order in $q$ and regarding the difference from $\hat\Pi^0(k,k',0)$ as a small departure, we then solve the problem by perturbation method. Since expanding $\hat\Pi^0(k,k',q)$ [equivalent to expanding $\hat h(k',q)$] is an elementary manipulation but tedious [because there are 16 elements in $\hat h(k',q)$], we here just present the result. For $l \ne 0$, the difference between $\hat\Pi^l$ and $\hat\Pi^0$ comes from the intervalley scattering term in $\hat W^l$. Similarly, we can treat this difference by perturbation. For all the cases, to the first order in the perturbation, we have 
\begin{eqnarray}
\lambda^l(q) &\approx &\frac{1}{\langle\Psi|\Psi\rangle V^2}\sum_{kk'}\Psi^{\dagger}(k)\hat\Pi^l(k,k',q)\Psi(k') \nonumber\\
&\approx &\lambda^l(0) - d_lq^2, ~~~~~~~~~~~~{\rm for}~~q\to 0 \label{lmbd}
\end{eqnarray}
where $\langle\Psi|\Psi\rangle$ = $\sum_k\Psi^{\dagger}(k)\Psi(k)/V$, $\lambda^l(0)$ and $d_l$ are positive constants. To the 0th order, the eigenfunction $\Psi(k)$ is unchanged. We here consider only the case of small $q$ since that is where QIC is significant.  

\begin{figure} 
\centerline{\epsfig{file=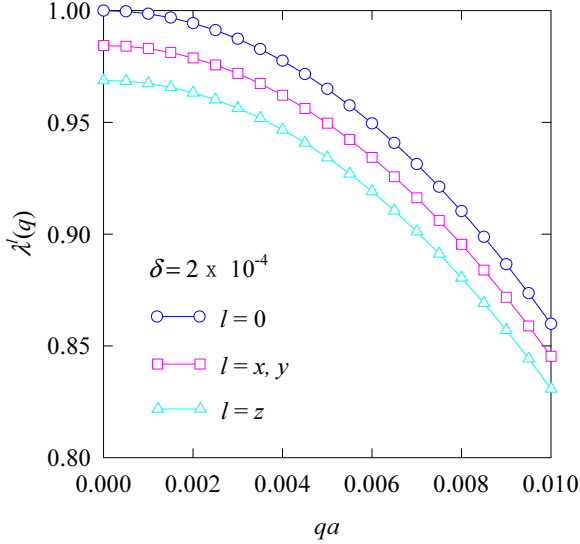,width=8. cm}}
\caption{(color online) Eigenvalue $\lambda^l(q)$ as function of $q$ at electron doping concentration $\delta = 2.0\times 10^{-4}$.}
\end{figure} 

In Fig. 3, the eigenvalues $\lambda^l(q)$ are shown as functions of $q$ at electron doping concentration $\delta = 2.0\times 10^{-4}$ (the doped electrons per site). The impurity scattering potential is given by the screened Coulomb one and the impurity concentration is chosen as  $n_i = 1.15\times 10^{-3}a^{-2}$ (with $a$ the lattice constant).\cite{Yan} The eigenvalue for $l = x,y$ is degenerated. In the limit $q \to$ 0, only $1-\lambda^0(q)$ approaches zero. The finite value $1-\lambda^l(0)$ for $l \ne 0$ is determined by the intervalley scattering strength $v_1$. At $v_1 = 0$, $\lambda^l(0) = 1$ is valid for all $l$. $1-\lambda^l(0)$ is larger for stronger $v_1$.

The state for each $l$ so obtained is of the lowest $|1-\lambda^l(q)|$. For the lowest state, $c^l_n(q)$ is given by $c^l(q) = c_l/[1-\lambda^l(q)]$ with 
\begin{eqnarray}
c_l =\frac{1}{\langle\Psi|\Psi\rangle^2 V^3}\sum_{kk_1k'}\Psi^{\dagger}(k)\hat\Pi^l(k,k_1,0)\hat W^l(|k_1-k'|)\Psi(k') \nonumber
\end{eqnarray}
where the $q$-dependence of $c_l$ has been neglected because of the drastic behavior of $1/[1-\lambda^l(q)]$ at small $q$. The Cooperon is finally approximated as
\begin{eqnarray}
C^l(k,k',q) = c_l\Psi(k)\Psi^{\dagger}(k')/[1-\lambda^l(q)]. \label{acp}
\end{eqnarray} 
For the zero-range scatters, only the isospin-singlet $C^l_{00}(k,k',q)$ survives and is independent of $k$ and $k'$. In this case with $v_1 = 0$ and $\omega = 0$, by the one-band approximation, one obtains $\lambda^l(q) = 1-Dq^2/4\Delta_0$ where $D = v^2/2\Delta_0$ is the diffusion constant and $\Delta_0$ is the first component of $\Psi$. The resultant Cooperon is $4\Delta_0n_iv_0^2/Dq^2$ consistent with that of Ref. \onlinecite{McCann} to the order of $q^{-2}$ in $q \to 0$. 

With the Cooperon $C^l$, the QIC to the current-current correlation function $\delta P(\omega^-,\omega^+)$ is calculated according to Fig. 2(b). Because the vertex, the Green functions and the Cooperons are matrices, one cannot write out $\delta P(\omega^-,\omega^+)$ immediately. For doing it, we start to work in the valley-sublattice space. According to the Feynman rule, we have,
\begin{widetext}
\begin{eqnarray}
\delta P(\omega^-,\omega^+)= \frac{2v^2}{V^2}\sum_{k q j \alpha} [V_x(-k,\omega^-,\omega^+)]^{j_4j_1}_{\alpha_4\alpha_1} C^{j_1j_2j_3j_4}_{\alpha_1\alpha_2\alpha_3\alpha_4}(-k,k,q,\omega)[V_x(k,\omega^+,\omega^-)]^{j_3j_2}_{\alpha_3\alpha_2} \label{ccc}
\end{eqnarray}
\end{widetext}
where $V_x(k,\omega_1,\omega_2) = G(k,\omega_1)\Gamma_x(k,\omega_1,\omega_2)G(k,\omega_2)$ is the vertex connected with two Green functions. With the inverse transform of Eq. (\ref{vlpi})
\begin{equation}
C^{j_1j_2j_3j_4}_{\alpha_1\alpha_2\alpha_3\alpha_4} =
\frac{1}{4}\sum_{ll'ss'}(M^{l\dagger}_{s})^{j_2j_1}_{\alpha_2\alpha_1}C^{ll'}_{ss'}(M^{l'}_{s'})^{j_3j_4}_{\alpha_3\alpha_4}, \nonumber
\end{equation}
we obtain
\begin{eqnarray}
\delta P(\omega^-,\omega^+)= \frac{v^2}{2V^2}\sum_{k q l}{\rm Tr}[Z^l(k,\omega)C^l(-k,k,q)] \label{P}
\end{eqnarray}
where $Z^l(k,\omega)$ is a matrix with elements $Z^l_{ss'}$ defined as 
\begin{equation}
Z^l_{ss'}= {\rm Tr}[V_x^t(k,\omega^+,\omega^-) M^l_sV_x(-k,\omega^-,\omega^+)M^{l\ast}_{s'}]. \nonumber 
\end{equation}
There is a simple relation, $Z^l(k,\omega) = -Z^0(k,\omega)$ for $l \ne 0$, because
\begin{eqnarray}
M^l_s = M^0_s\Lambda_l,~~~~\Lambda_lM^0_s\Lambda^{\ast}_l = -M^0_s,~~~~{\rm for}~~l = x,y,z \nonumber
\end{eqnarray}
and the operator $\Lambda_l$ commutes with $G$ and $\Gamma_x$. This result means that the QIC by the pseudospin singlet ($l = 0$) is negative, while it is positive by the pseudospin triplets ($l = x,y,z$). Substituting the results given by Eqs. (\ref{lmbd}) and (\ref{acp}) into Eq. (\ref{P}) and carrying out the $q$-integral, we get 
\begin{eqnarray}
\delta P(\omega^-,\omega^+) = f\sum_lN_l\frac{c_l}{d_l}\ln\frac{1-\lambda^l(0)+d_lq^2_1}{1-\lambda^l(0)+d_lq^2_0},\label{P1}\\
f = \frac{v^2}{8\pi V}\sum_{k}\Psi^{\dagger}(k)Z^0(k,\omega)\Psi(-k),
\end{eqnarray}
where $N_0 = -1$, $N_{l=x,y,z} = 1$, $q_0$ and $q_1$ are the lower and upper cutoffs of the $q$-integral. 

The lower cutoff $q_0$ is given by $q_0 = {\rm max}(L^{-1}_{in},L^{-1})$ where $L_{in}$ is the length the electrons diffuse within an inelastic collision time $\tau_{in}$ and $L$ the length scale of the system.\cite{Lee} At very low doping ($\delta < 1.0\times10^{-3}$, the doped electrons per site), $\tau_{in}$ due to the inter-electronic Coulomb interaction is estimated as $\tau_{in} \approx 0.462v/aT^2$ (where $T$ is the temperature) from the recent study of the interacting electrons in graphene using renormalized-ring-diagram approximation.\cite{Yan1} $L_{in}$ is then given by $L_{in} = (v^2\tau\tau_{in}/2)^{1/2}$ where the elastic collision time $\tau$ is determined by the non QIC-corrected conductivity $\sigma_0$, $\tau = \hbar\pi \sigma_0/vk_Fe^2$ (with $k_F$ as the Fermi wavenumber).\cite{Lee} For low carrier density, we find that $L_{in}$ is about a few microns for 4 K $< T <$ 20 K. On the other hand, the upper limit is $q_1 = L_0^{-1}$ with $L_0 = v\tau$ as the length of mean free path.

Using the present formalism, we have recently studied the WL effect of Dirac fermions in graphene.\cite{Yan2} It is found that WL is present in large size samples at finite carrier doping. The strength of WL becomes weakened/quenched when the sample size $< L_{in}$ (about a few microns at low temperatures) as studied in the experiment.\cite{Geim,Morozov} Close to region of zero doping, the system may be delocalized. Physically, at small electron doping, the Fermi circle and the typical momentum transfer $q$ ($\sim 2k_F$) are small and also the screening is weak, leading to stronger $v_0(q)$ than $v_1$. For weak intervalley scattering, all $\lambda^l(0)$'s ($l \ne 0$) close to 1, the QIC from each pseudospin channel has almost the same magnitude. After one of $l \ne 0$ is canceled by the $l = 0$ channel, the net QIC is positive. This is consistent with the fact that Dirac fermions cannot be scattered to exactly the backwards direction in case of $v_1 = 0$ and the WL is absent. On the other hand, with increasing electron doping, the strength of the intervalley scatterings becomes stronger, leading to the appearance of WL in large size samples. The detailed numerical study of the WL in graphene has been presented in Ref. \onlinecite{Yan2} and will not be repeated here.

We here concisely explain why pseudospin singlet Cooperons give rise to WL (negative QIC) but the pseudospin triplets result in anti-WL (positive QIC). As seen from the matrix $M^l_s$ defined below Eq. (\ref{vlpi}), the pseudospins are actually associated with the matrices $\Lambda_y\Lambda_l$, 
\begin{eqnarray}
\Lambda_y\Lambda_0 &=& \tau_2\sigma_3,~~~{\rm singlet}\nonumber\\ 
\Lambda_y\Lambda_x &=& -i\tau_3,~~~{\rm triplet~with}~l = x \nonumber\\
\Lambda_y\Lambda_y &=& 1,~~~~~~~{\rm triplet~with}~l = y \nonumber\\
\Lambda_y\Lambda_z &=& i\tau_1\sigma_3, ~~{\rm triplet~with}~l = z. \nonumber 
\end{eqnarray}
In terms of the $c$-operator, $\psi^t_{q/2-k}\Lambda_y\Lambda_l\psi_{q/2+k}$ annihilates a Cooperon of total momentum $q$ and relative momentum $k$ with pseudospin $l$. It is seen that in the pseudospin singlet state the two particles are in different valleys and the parity is odd under the exchange of the valley indices. The magnitude of the wave function of the pseudospin singlet Cooperon is large when the two particles occupy respectively the opposite momentum [defined respect to the origin of the Brillouin zone, see Fig. 1(a)] states of the single particles. It implies a strong backward scattering for the electrons and thereby leads to WL. On the other hand, the pseudospin triplets are even under the valley exchange. For $l = z$, though the two particles are in different valleys, the wave function of the Cooperon is small when the two particles occupy respectively the opposite momentum states. In this case, the backward scattering is weakened, resulting in the increase of the conductivity. For $l = x$ and $y$, the two particles occupy the states in the same valley and their total momentum is finite. The case corresponds to the final state of the scattered electrons is not in the backward direction, giving rise to a positive contribution to the conductivity. The anti-WL can be reduced when the backward scattering is strengthened by the intervalley scatterings.  

\section{Magnetoconductivity}

\begin{figure} 
\centerline{\epsfig{file=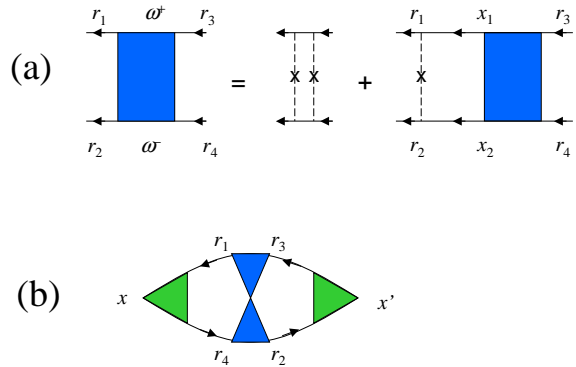,width=8. cm}}
\caption{(color online) (a) Cooperon propagator in real space. (b) Quantum interference correction to the conductivity.}
\end{figure} 

We here consider that the system is acted with an external magnetic field $B$ perpendicular to the graphene plane. The WL of Dirac fermions in the existence of weak magnetic field can be treated in a way parallel to Ref. \onlinecite{Altshuler}. Here, we outline the main steps. Since the system becomes an inhomogeneous one in this case, we need to start with the description in real space. The kinetic part of the Hamiltonian $H_0$ is 
\begin{equation}
H_0 = \int d\vec r\tilde\psi^{\dagger}(r)v\tau_z\vec\sigma
 \cdot(\vec p + \vec A)\tilde\psi(r) \label{Hm}
\end{equation}
where $\vec p$ is the momentum operator, and $\vec A$ is the vector potential. Here, we have used the units of $e = c = 1$ (with $-e$ as an electron charge and $c$ the light velocity). With gauge transform \begin{equation}
\tilde\psi(r) = \psi(r)\exp(-i\int^r_{r_0}d\vec l \cdot\vec A), \nonumber
\end{equation}
$\vec A$ is eliminated from $H_0$. On viewing this gauge transform, for very weak magnetic field $B$, the Green function $\tilde G(r,r',\tau-\tau') = -\langle T\psi(r,\tau)\psi^{\dagger}(r',\tau')\rangle$ can be approximated as\cite{Altshuler}
\begin{equation}
\tilde G(r,r',\omega) \approx G(r-r',\omega)\exp(i\int^{r'}_{r}d\vec l\cdot \vec A), \label{gf}
\end{equation}
where $G(r-r',\omega)$ is the Green function in the absence of the magnetic field. Here, the position $r$ is understood as vector.

In real space, the Bethe-Salpeter equation for the Cooperon is diagrammatically shown in Fig. 4(a). One can then write out it explicitly and do the same transform as from Eq. (\ref{cp0}) to Eq. (\ref{bsh0}). The final matrix equation (in the isospin space) is given by 
\begin{widetext}
\begin{eqnarray}
\hat C^{l}(r_1,r_2,r_3,r_4) &=& \hat W^l(r_1-r_2)[\hat h(r_1,r_2,r_3,r_4)\hat W^l(r_3-r_4) 
+\int dx_1\int dx_2\hat h(r_1,r_2,x_1,x_2)\hat C^l(x_1,x_2,r_3,r_4)]  \label{cpr}
\end{eqnarray}
where $\hat W^l(r)$ is the real space representation of $\hat W^l(q)$, and the element of $\hat h(r_1,r_2,r_3,r_4)$ is given by
\begin{eqnarray}
h_{ss'}(r_1,r_2,r_3,r_4) = \frac{1}{4}{\rm Tr}[\tilde G(r_1,r_3,\omega^+)\Sigma_s\tilde G(r_2,r_4,\omega^-)\Sigma^{\dagger}_{s'}].\nonumber
\end{eqnarray}
Using the approximation given by Eq. (\ref{gf}), we have
\begin{eqnarray}
h_{ss'}(r_1,r_2,r_3,r_4) &\approx & h^0_{ss'}(r_1-r_3,r_2-r_4)\exp(i\int_{r_1}^{r_3}d\vec l\cdot\vec A+i\int_{r_2}^{r_4}d\vec l\cdot\vec A)   \nonumber\\
&\approx & h^0_{ss'}(R-R'+\frac{r-r'}{2},R-R'-\frac{r-r'}{2})\exp(i2\int_{R}^{R'}d\vec l\cdot\vec A) \label{hr}
\end{eqnarray}
where $R = (r_1+r_2)/2$, $r = r_1-r_2$,  $R' = (r_3+r_4)/2$, $r' = r_3-r_4$, and $h^0$ is defined in the absence of $\vec A$. The Fourier transform of $\hat h^0(R-R'+\frac{r-r'}{2},R-R'-\frac{r-r'}{2})$ is given by
\begin{eqnarray}
\hat h^0(R-R'+\frac{r-r'}{2},R-R'-\frac{r-r'}{2}) = \frac{1}{V^2}\sum_{kq}\hat h(k,q)e^{i\vec q\cdot(\vec R-\vec R')+i\vec k\cdot(\vec r-\vec r')}
\end{eqnarray}
where $h(k,q)$ is defined by Eq. (\ref{hkq}). The eigenvalue problem of Eq. (\ref{cpr}) reads
\begin{eqnarray}
\hat W^l(r_1-r_2)\int dr'_1\int dr'_2\hat h(r_1,r_2,r'_1,r'_2)\psi(r'_1,r'_2) = E^l\psi(r_1,r_2).  \label{engr}
\end{eqnarray}
where $\psi(r_1,r_2)$ is a four-component vector in the isospin space, and $E^l$ is the eigenvalue. Using the coordinates $R$ and $r$, we separate $\psi(r_1,r_2)$ as $\psi(r_1,r_2) = \Phi(R)\Psi(r)$
where $\Phi(R)$ is a scalar representing the motion of center of mass, and $\Psi(r)$ is a four-component vector meaning the relative motion of the Cooperon. Since the magnetic field is weak, only the large-scale motion of the center of mass is significantly affected; the magnetic filed influence on the relative motion is negligible. Then, $\Psi(r)$ can be considered as the real space representation of $\Psi(k)$ given in Sec. II. By integrating out the relative motion, we get
\begin{eqnarray}
\frac{1}{V}\sum_{q}\int dR'\lambda^l(q)\exp[i\vec q\cdot(\vec R-\vec R')+i2\int_R^{R'}d\vec l\cdot \vec A]\Phi(R') = E^l\Phi(R)  \label{cm}
\end{eqnarray}
\end{widetext}
with $\lambda^l(q)$ defined by Eq. (\ref{lmbd}). Using $\lambda^l(q)\approx \lambda^l(0) - d_lq^2$ for small $q$, $\Phi(R') = \exp[(\vec R'-\vec R)\cdot\nabla]\Phi(R)$, and carrying out the $q$-summation and $R'$-integral, we obtain
\begin{eqnarray}
[\lambda^l(0)-d_l(\vec P+2\vec A)^2]\Phi(R) = E^l\Phi(R),  \label{cms}
\end{eqnarray}
where $\vec P = -i\nabla$ is the momentum operator (of the center of mass) of the Cooperon. Using the Landau gauge $\vec A = (0,Bx,0)$, one has
\begin{eqnarray}
E^l_n = \lambda^l(0) - 4d_lB(n+1/2), \label{LD}
\end{eqnarray}
and $\Phi_n(R)$ being the wavefunction of the corresponding Landau state. The degeneracy of each level is $g = BV/\pi$. The Cooperon is obtained as
\begin{eqnarray}
\hat C^{l}(r_1,r_2,r_3,r_4) = g\sum_n c_l\frac{\Phi_n(R)\Phi_n^{\dagger}(R')}{1-E^l_n}\Psi(r)\Psi^{\dagger}(r'). \nonumber
\end{eqnarray}

The QIC to the conductivity in the presence of magnetic field is calculated according to Fig. 4(b). To explicitly write out the expression, we note that the magnetic field effect on the vertex is negligible small. The strong dependence of $B$ in QIC comes from the Cooperon due to the small denominator $1-E^l_n$. Therefore, the vertex associated with $x'$ in Fig. 4(b) connected with two Green functions after integrating over $x'$ is given by
\begin{eqnarray}
V_{x'}(r_3-r_2,\omega^+,\omega^-) = \frac{1}{V}\sum_{k_1}V_{x}(k_1,\omega^+,\omega^-)e^{i\vec k_1\cdot(\vec r_3-\vec r_2)} \nonumber
\end{eqnarray}
where $V_{x}(k_1,\omega^+,\omega^-)$ has appeared above Eq. (\ref{P}). Similarly, for the vertex associated with $x$ in Fig. 4(b), one gets
\begin{eqnarray}
V_{x}(r_4-r_1,\omega^-,\omega^+) = \frac{1}{V}\sum_{k_2}V_{x}(k_2,\omega^-,\omega^+)e^{i\vec k_2\cdot(\vec r_4-\vec r_1)}. \nonumber
\end{eqnarray}
The QIC to the current-current correlation function is 
\begin{widetext}
\begin{eqnarray}
\delta P(\omega^-,\omega^+)= \frac{v^2}{2V}\sum_{l}\int dr_1\cdots\int dr_4{\rm Tr}[Z^l(r_3-r_2,r_4-r_1,\omega)C^l(r_1,r_2,r_3,r_4)] \label{Pr}
\end{eqnarray}
with $Z^l_{ss'}(r_3-r_2,r_4-r_1,\omega)= {\rm Tr}[V_{x'}^t(r_3-r_2,\omega^+,\omega^-) M^l_sV_x(r_4-r_1,\omega^-,\omega^+)M^{l\ast}_{s'}]$. Substituting the results for $V_{x'}$, $V_{x}$, and $C^l(r_1,r_2,r_3,r_4)$ into Eq. (\ref{Pr}), setting $k_{1,2} = \pm k+q/2$ and neglecting the small $q$-dependence in $V_{x}(\pm k+q/2,\omega_1,\omega_2)$, then using the coordinates $R$, $R'$, $r$, and $r'$, and integrating out the relative motions, we get
\begin{eqnarray}
\delta P(\omega^-,\omega^+) = \frac{4\pi gf}{V^2}\sum_{lnq}\frac{N_lc_l}{1-E^l_n}\int dR\int dR'\Phi_n(R)\Phi_n^{\dagger}(R')e^{i\vec q\cdot(\vec R'-\vec R)} 
= \frac{4\pi gf}{V}\sum_{ln}\frac{N_lc_l}{1-E^l_n}. \label{pcr}
\end{eqnarray}
\end{widetext}
By comparing this result with Eq. (\ref{P}), we see that the $q$-integral in Eq. (\ref{P}) is replaced with the summation over the Landau levels. By using the same $q$-cutoffs as in obtaining Eq. (\ref{P1}), only those states of energy levels $\epsilon^l_n =B(2n+1)$ (with $E^l_n \equiv \lambda^l(0) - 2d_l\epsilon^l_n$) in the range $(q^2_0/2, q^2_1/2)$ (in units of $a$ = $v$ = 1) need to be summed up. 

With the current-current correlation function, one then calculates the conductivity $\sigma$ according to the Kubo formalism.\cite{Mahan} At very low temperatures, the correction to the conductivity is calculated by
\begin{eqnarray}
\delta\sigma = \delta P(0^-,0^+)/2\pi,
\end{eqnarray}
which depends on the magnetic field. The magnetoconductivity is defined as $\Delta\sigma(B) = \sigma(B)-\sigma(0)$ where $\sigma$ is the corrected conductivity including the non-corrected one (see Ref. \onlinecite{Yan}) and the correction $\delta\sigma$.

Shown in Fig. 5 are the present results for the magnetoconductivity $\Delta\sigma(B)$ and comparison with experiments. For comparing with the experiments, we note that the overall magnitude of $\Delta\sigma(B)$ varies largely from sample to sample (of the same carrier density) and from experiment to experiment.\cite{Ki,Tikhonenko} Instead to analyzing this variation, we here confine ourselves to see the magnetic field dependence of $\Delta\sigma(B)$ and therefore depict the results for the normalized magnetoconductivity. The solid, dashed, and dotted lines in Fig. 5 are the present calculations for the parameters ($T, \delta, L$) = (0.12 K, $8\times 10^{-4}$, 1$\mu$m), (0.12 K, $1.5\times 10^{-3}$, 1$\mu$m), and (7 K, $2\times 10^{-4}$, 2$\mu$m), respectively. These sets of parameters are close to the conditions for the two experiments: the red filled circles and green filled squares from Ref. \onlinecite{Ki}, the up and down triangles and the diamonds from Ref. \onlinecite{Tikhonenko}. The impurity density $n_i = 1.15\times 10^{-3}a^{-2}$ and the scattering potential used here are the same as in the previous work\cite{Yan} that reproduces the zero-field electric conductivity of the experimental results.\cite{Geim} Clearly, the present calculation is in good agreement with the experimental measurements. Notice that there is no adjustable free parameter in the present calculation.

The magnetoconductivity comes from the pseudospin triplet channels of the Cooperon. From Eq. (\ref{pcr}), we see that the contribution $\delta P_{nl}(B)$ to the conductivity from each Landau level of the center of mass of the Cooperon is,
\begin{equation}
\delta P_{nl}(B) \propto \frac{B}{1-\lambda^l(0)+2d_l(2n+1)B}.
\end{equation}
Since $\lambda^0(0) = 1$, $\delta P_{n0}$ is independent of $B$ and there is no contribution from the pseudospin singlet channel to $\Delta\sigma(B)$ (both constant terms respectively in $\sigma(B)$ and $\sigma(0)$ cancel each other). For the pseudospin triplet, $1-\lambda^l(0) > 0$, and at small $B << [1-\lambda^l(0)]/2d_l(2n+1)$, $\delta P_{nl}(B)$ increases linearly with $B$. But the total contribution to $\delta\sigma(B)$ does not vanish at $B = 0$ because the Landau levels become continuum and the $q$-integral is restored giving rise to a constant. The total contribution $\delta\sigma(B)$ at $B = 0$ again cancel with the corresponding term in $\sigma(0)$. The final result for $\Delta\sigma(B)$ depends delicately on $B$. The function $\Delta\sigma(B)$ at weak $B$ is determined by the constants $\lambda^l(0)$ and $d_l$ for $l \ne 0$. From Fig. 3, we have $[1-\lambda^z(0)]/d_z \approx 2[1-\lambda^l(0)]/d_l|_{l=x,y} \equiv \tau^{-1}_s$. Therefore, the behavior of $\Delta\sigma(B)$ can be characterized by the constant $\tau^{-1}_s$. The quantity $\tau^{-1}_s$ is a measure of the intervalley scattering [because of $1-\lambda^z(0)$ as mentioned in Sec. II].

\begin{figure} 
\centerline{\epsfig{file=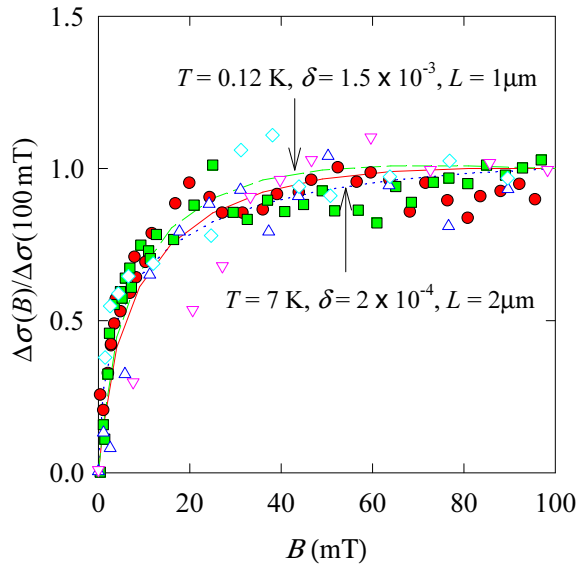,width=8. cm}}
\caption{(color online) Normalized magnetoconductivity as function of the magnetic field $B$. The lines are the present calculations: solid, dashed, and dotted lines are for the parameter sets ($T, \delta, L$) = (0.12 K, $8\times 10^{-4}$, 1$\mu$m), (0.12 K, $1.5\times 10^{-3}$, 1$\mu$m), and (7 K, $2\times 10^{-4}$, 2$\mu$m), respectively. The symbols are the experimental results: red filled circles ($\delta = 8\times 10^{-4}$ corresponding to $n = 3.15\times 10^{12}$ cm$^{-2}$) and green filled squares [$\delta = 1.5\times 10^{-3}$ ($n = 6.1\times 10^{12}$ cm$^{-2}$)], all at $T$ = 0.12 K and $L \approx 1 \mu$m, are from Ref. \onlinecite{Ki}; up triangles [$T$ = 7 K, $\delta = 2\times 10^{-4}$ ($n = 8\times 10^{11}$ cm$^{-2}$)], down triangles ($T$ = 4 K, $\delta = 2\times 10^{-4}$), and diamonds [$T$ = 0.26 K, $\delta = 2.5\times 10^{-4}$ ($n = 10^{12}$ cm$^{-2}$)] all with $L \approx 2 \mu$m are from Ref. \onlinecite{Tikhonenko}. The magnetic field is given in unit of $10^{-3}$ Tesla.}
\end{figure}

\section{Summary}

In summary, on the basis of self-consistent Born approximation, we have presented the WL theory of the Dirac fermions under the charged impurity scatterings in graphene. The Bethe-Salpeter matrix equations for the Cooperons are solved by perturbation method. There are three non-vanishing components in the wavefunctions of Cooperons under the finite-range impurity scatterings. This is different from the zero-range one. The pseudospin singlet and triplet Cooperons give rise to WL and anti-WL effect, respectively. For small carrier doping where the intervalley scatterings are much weaker than the intravalley scatterings, the anti-WL effect is predominant over the WL one. While for large carrier doping, the WL effect is significant because of the intravalley scatterings weakened. The WL effect is also determined by the sample size. It is found that WL is quenched at low temperature when the sample size is smaller than the inelastic collision length. The latter is about a few microns for 4 K $< T <$ 20 K and at low carrier concentrations.\cite{Yan2} With the model of charged impurity scatters justified for the electric conductivity at zero-magnetic field, we have calculated the magnetoconductivity. For weak external magnetic field, the magnetoconductivity comes from the contributions of pseudospin triplet Cooperons. It is shown that the present results for the magnetoconductivity are in good agreement with some of the experimental measurements.

\acknowledgments

This work was supported by a grant from the Robert A. Welch Foundation under No. E-1146, the TCSUH, the National Basic Research 973 Program of China under grant No. 2005CB623602, and NSFC under grant No. 10774171 and No. 10834011.

\vskip 5mm
\centerline {\bf APPENDIX}
\vskip 3mm

In this appendix, we discuss the impurity scattering potential. In our previous work,\cite{Yan} we have illustrated how the intravalley and intervalley scatterings $v_0(q)$ and $v_1$ in Eq. (\ref{vi}) within the SCBA are determined from the microscopic electron-impurity interactions. In our numerical calculations, $v_0(q)$ and $v_1$ are set as respectively the values of leading terms in their expansions for the charged impurities. In the derivation, the difference of the $a$ and $b$ sites in the same unit cell of graphene lattice was neglected. As long as the long-wave length scatterings are considered, such a difference is negligible. It is true for the intravalley scatterings where the long-wave length scatterings are predominant momentum transfers of the Dirac fermions. While for the intervalley scatterings, the momentum transfers are finite and more carefulness are needed. Here, taking into account the sublattice difference, we show that only the leading terms of $v_0(q)$ and $v_1$ need to be included in the effective potentials. 

\begin{figure} 
\centerline{\epsfig{file=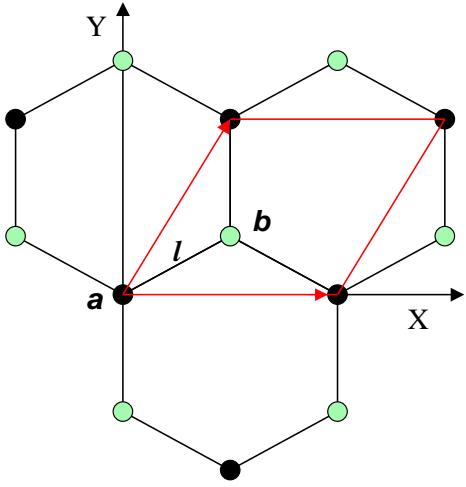,width=8. cm}}
\caption{(color online) The structure of a honeycomb lattice. There are two sites, $a$ (black) and $b$ (green), in each unit cell enclosed by the red lines. {\it l} is a vector from $a$ to $b$.} 
\end{figure}

We start with the Hamiltonian for electron-impurity interactions in graphene,
\begin{eqnarray}
H_1 &=& \sum_{j\alpha}\int d\vec Rn_{\alpha}(\vec r_j)v_{\alpha i}(|\vec
 r_j-\vec R|)n_i(\vec R)  \nonumber\\
&=& \frac{1}{V}\sum_{q\alpha}n_{\alpha}(q)v_{\alpha i}(q)n_i(-q) , \label{H1}
\end{eqnarray}
where $n_{\alpha}(\vec r_j)$ is the density operator of electrons at $\alpha$ (= $a$ or $b$) cite of $j$th unit cell of the honeycomb lattice (Fig. 6), $n_i(\vec R)$ is the real space density distribution of impurities, and $v_{\alpha i}(|\vec r_j-\vec R|)$ is the impurity scattering potential. The Fourier component of the electron density $n_{\alpha}(q) = \sum_kc^{\dagger}_{k-q\alpha}c_{k\alpha}$ can be written as  
\begin{eqnarray}
n_{\alpha}(q) \approx \sum_k{'} (c^{\dagger}_{k-q\alpha 1}c_{k\alpha 1}+c^{\dagger}_{k-q\alpha 2}c_{k\alpha 2}) ~~~~{\rm for}~q\sim 0 \nonumber \\
n_{\alpha}(q) \approx 
\begin{cases}
\sum\limits_k{'}c^{\dagger}_{k-q'\alpha 2}c_{k\alpha 1} ~~~~{\rm for}~q\sim Q+q'\sim Q  \\ 
\sum\limits_k{'}c^{\dagger}_{k-q'\alpha 1}c_{k\alpha 2} ~~~~{\rm for}~q\sim -Q+q'\sim -Q 
\end{cases}\nonumber
\end{eqnarray}
where $k$ summation $\sum'$ runs over a valley since we here consider only the low energy excitations. In the main text, the $k$ summation $\sum$ means $\sum'$, and we hereafter use the simple notation $\sum$. In Eq. (\ref{H1}), the $q$ summation runs over the infinitive momentum space. By folding the whole space into the first Brillouin zone, $H_1$ can be written as 
\begin{eqnarray}
H_1 = \frac{1}{V}\sum_{kq}\psi^{\dagger}_{k-q}V_i(q)\psi_{k} , \label{H2}
\end{eqnarray}
where $k,q$ run over a valley in the Brillouin zone, and 
\begin{equation}
V_i(q) = 
\begin{pmatrix}
V_d(q)& V_o(q-Q) \\
V_o^t(q+Q)& \tilde V_d(q)
\end{pmatrix}\label{v1}
\end{equation}
with
\begin{eqnarray}
V_d(q) &=& 
\begin{pmatrix}
\phi_a(q)& 0 \\
0& \phi_b(q)
\end{pmatrix}, ~~~~
\tilde V_d(q) = 
\begin{pmatrix}
\phi_b(q)& 0 \\
0& \phi_a(q)
\end{pmatrix} \nonumber\\
V_o(q\pm Q) &=& 
\begin{pmatrix}
0& \phi_a(q\pm Q)_{atc} \\
 \phi_b(q\pm Q)_{atc}&0
\end{pmatrix} \nonumber\\
\phi_a(q)&=& \sum_nn_i(-q-Q_n)V(q+Q_n)\nonumber\\
\phi_b(q)&=& \sum_ne^{i(\vec q+\vec Q_n)\cdot\vec l}n_i(-q-Q_n)V(q+Q_n). \nonumber
\end{eqnarray}
Here $V(q)$ is the Fourier component of $V_{ai}(r)$, and $Q_n$ is the reciprocal lattice vector.  There is a phase factor $e^{i(\vec q+\vec Q_n)\cdot\vec l}$ in the expansion of $\phi_b(q)$ because of the position difference $l$ between $a$ and $b$ sites. The expressions for $\phi_{a,b}(q)_{atc}$ are similar to $\phi_{a,b}(q)$ except avoiding triple counting since $\phi_{a,b}(q\pm Q)_{atc}$ imply the intervalley scatterings. [There are other two equivalent valleys for each valley indicated in Fig. 1(a).] For example, the leading order in $\phi_{a}(q-Q)_{atc} \approx \phi_{a}(-Q)_{atc}$ is $n_i(\overline Q)V(\overline Q)$ where $\overline Q$ (with $|\overline{Q}| = 4\pi/3a$, $a \sim$ 2.4 \AA~ as the lattice constant) is a  momentum difference between the nearest-neighbor Dirac points in the Brillouin zone. For this $\overline Q$, there are other two vectors of the same magnitude differ from $\overline Q$ by two reciprocal lattice vectors [see Fig. 1(a)], respectively. These two terms should be excluded from the summation. Since the case of $q \sim 0$ is under consideration, we approximate the off-diagonal potential as $V_o(q\pm Q) \approx V_o(\pm Q)$. 

We have argued\cite{Yan} that if the effective impurity scatterings are given as Eq. (\ref{vi}), then the vertex $\Gamma_x(\vec k,\omega_1,\omega_2)$ can be expanded as Eq. (\ref{vt}). [The off-diagonal parts in Eq. (\ref{vi}) are different from that in Ref. \onlinecite{Yan} where $\sigma_0$ was used instead of $\sigma_1$ because the basis was  $\psi^{\dagger}_{k}=(c^{\dagger}_{ka1},c^{\dagger}_{kb1},c^{\dagger}_{ka2},c^{\dagger}_{kb2})$ with reflected $y$-axis in valley 2. But the result for $\Gamma_x(\vec k,\omega_1,\omega_2)$ is unchanged.] Here, suppose the vertex $\Gamma_x(\vec k,\omega_1,\omega_2)$ is given as Eq. (\ref{vt}), we want to see how $v_0(q)$ and $v_1$ in Eq. (\ref{vi} are expected from the microscopic potentials given by Eq. (\ref{v1}). The $x$-direction current vertex $v\Gamma_x(\vec k,\omega_1,\omega_2)$ under the SCBA satisfies the following $4\times 4$ matrix integral equation,
\begin{widetext}
\begin{equation}
\Gamma_x(\vec k,\omega_1,\omega_2)=
 \tau_3\sigma_x+\frac{1}{V^2}\sum_{k'}\langle V_i(\vec k-\vec k')G(\vec k',\omega_1)\Gamma_x(\vec
 k',\omega_1,\omega_2)G(\vec k',\omega_2)V_i(\vec k'-\vec k)\rangle,\label{vt1}
\end{equation}
where $\langle\cdots\rangle$ means the average over the impurity distributions [Fig. 1(c)]. Notice that the product $G(\vec k,\omega_1)\Gamma_x(\vec  k,\omega_1,\omega_2)G(\vec k,\omega_2)$ can be expanded in $A^x_j(\hat k)$,
\begin{equation}
G(\vec k,\omega_1)\Gamma_x(\vec  k,\omega_1,\omega_2)G(\vec k,\omega_2) = \sum_{jj'}A^x_j(\hat k)L_{jj'}(\vec k,\omega_1,\omega_2)y_{j'}(\vec  k,\omega_1,\omega_2), \nonumber
\end{equation}
where the functions $L_{jj'}(\vec k,\omega_1,\omega_2)$ have been defined in Ref. \onlinecite{Yan}. We then need to calculate the expectations $\langle V_i(\vec k-\vec k')A^x_j(\hat k')V_i(\vec k'-\vec k)\rangle$ in Eq. (\ref{vt1}). Firstly, we calculate $\langle V_i(q)A^x_0V_i(-q)\rangle = A^x_0[\langle\phi_a(q)\phi_b(-q)\rangle-\langle\phi_a(Q)_{atc}\phi_b(-Q)_{atc}\rangle$. Using the expressions for $\phi_{a,b}(q)$, we obtain
\begin{eqnarray}
\langle\phi_a(q)\phi_b(-q)\rangle
= Vn_i\sum_nV^2(q+Q_n)e^{-i(\vec q+ \vec Q_n)\cdot \vec l} \approx Vn_iV^2(q), \nonumber
\end{eqnarray}
where the use of the fact that $\sum_{n\ne 0}|V(q+Q_n)|^2e^{-i\vec Q_n\cdot \vec l} = 0$ for $q \sim 0$ has been made. Similarly, we get
\begin{eqnarray}
\langle\phi_a(Q)_{atc}\phi_b(-Q)_{atc}\rangle 
= Vn_i\sum_n{'}V^2({\overline Q}+Q_n)e^{-i(\vec{\overline Q}+ \vec Q_n)\cdot \vec l}
\approx Vn_iV^2({\overline Q}), \nonumber
\end{eqnarray}
where $\sum_n{'}$ means avoiding triple counting, and $\vec{\overline Q}\cdot \vec l = 0$ because for $\vec l$ there is a $\vec{\overline Q}$ orthogonal to it [see Figs. 1(a) and 5]. For $A^x_j(\hat k')$ with $j \ne 0$, we note
\begin{eqnarray}
A^x_{1,2}(\hat k') = A^x_{1,2}(\hat k)(\cos\theta\pm i\sigma_z\sin\theta),~~
A^x_3(\hat k') = A^x_3(\hat k)(\cos 2\theta+i\sigma_z\sin 2\theta),
\end{eqnarray}
where $\theta$ is the angle between $\hat k$ and $\hat k'$. Since $V(\vec k'-\vec k)$ depends on $\theta$ through $\cos\theta$, the integrals of the integrands with factor $\sin\theta$ or $\sin 2\theta$ vanish. $\langle V_i(\vec k-\vec k')A^x_j(\hat k')V_i(\vec k'-\vec k)\rangle$ for $j \ne 0$ are then calculated as
\begin{eqnarray}
\langle V_i(\vec k-\vec k')A^x_{1,2}(\hat k')V_i(\vec k'-\vec k)\rangle \to A^x_{1,2}Vn_i\sum_nV^2(|\vec k-\vec k'+Q_n|)\cos\theta 
\approx A^x_{1,2}Vn_iV^2(|\vec k'-\vec k|)\cos\theta       \nonumber\\
\langle V_i(\vec k-\vec k')A^x_3(\hat k')V_i(\vec k'-\vec k)\rangle \to A^x_3Vn_i\sum_nV^2(|\vec k-\vec k'+Q_n|)e^{-i(\vec q+ \vec Q_n)\cdot \vec l}\cos 2\theta 
= A^x_3Vn_iV^2(|\vec k'-\vec k|)\cos 2\theta.       \nonumber
\end{eqnarray}
By defining $v_0^2(q) = V^2(q)$ and $v_1^2 = V^2(\overline Q)$, we obtain exactly the same equation as Eq. (11) in Ref. \onlinecite{Yan} for determining $y_j(\vec  k,\omega_1,\omega_2)$. Thus, we have proved that only the leading terms of $v_0(q)$ and $v_1$ are necessarily taken into account in the current vertex corrections. 

On the other hand, under the SCBA, the self-energy is given by
\begin{eqnarray}
\Sigma(\vec k,\omega)&=&
\frac{1}{V^2}\sum_{k'}\langle V_i(\vec k-\vec k')G(\vec k',\omega)V_i(\vec k'-\vec k)\rangle \nonumber\\
&\approx &\frac{n_i}{V}\sum_{k'}\{[\sum_nV^2(\vec k-\vec k'+Q_n)+\sum_n{'}V^2(Q-Q_n)]G_0(\vec k',\omega) + V^2(\vec k-\vec k')G_c(\vec k',\omega)\hat k\cdot\hat k'\hat k\cdot\vec\sigma\tau_z\}  \nonumber \\
&\approx &\frac{n_i}{V}\sum_{k'}\{[V^2(\vec k-\vec k')+ V^2(\overline Q)]G_0(\vec k',\omega) + V^2(\vec k-\vec k')G_c(\vec k',\omega)\hat k\cdot\hat k'\hat k\cdot\vec\sigma\tau_z\},  \nonumber 
\end{eqnarray}
\end{widetext}
which is consistent with Eqs. (\ref{sc2}) and (\ref{sc3}). Therefore, the theory is simplified with the effective potential given by Eq. (\ref{vi}) with $v_0(q)$ and $v_1$ defined above.

\end{document}